\documentclass{article}
\usepackage{ijcai26}

\usepackage{times}
\usepackage{soul}
\usepackage{url}
\usepackage[hidelinks]{hyperref}
\usepackage[utf8]{inputenc}
\usepackage[small]{caption}
\usepackage{graphicx}
\usepackage{amsmath}
\usepackage{amsthm}
\usepackage{booktabs}
\usepackage{algorithm}
\usepackage{algorithmic}
\usepackage[switch]{lineno}

\title{MC-RAG System: A Structure-Driven RAG System for Multi-Constraint Queries}

\author{
  Xiao Zhang$^1$
  \and
  Yang Wan$^1$
  \and
  Yi Li$^2$
  \and
  Miao Xie$^1$
  \and
  Chunli Lv$^{1,3}$\\
  \affiliations
  $^1$College of Information and Electrical Engineering, China Agricultural University, China\\
  $^2$College of Computing and Data Science, Nanyang Technological University, Singapore\\
  $^3$Key Laboratory of Agricultural Informatization Standardization, Ministry of Agriculture and Rural Affairs, China\\
  \emails
  B20243080802@cau.edu.cn,
  SY20253082208@cau.edu.cn,
  liyi0067@e.ntu.edu.sg,
  0520shui@163.com,
  lvcl@cau.edu.cn
}

\begin{document}

\maketitle

\begin{abstract}
Retrieval-Augmented Generation (RAG) systems are widely adopted in question answering, yet they often fail to satisfy complex multi-constraint queries, leading to constraint violations, factual inconsistencies, or hallucinations.
We present \textsc{ Structure-Driven RAG System for Multi-Constraint Queries(MC-RAG)}, 
a structure-driven RAG system that reformulates retrieval as a subgraph matching problem over a knowledge graph. By integrating semantic and structural embeddings with path-level indexing, \textsc{MC-RAG} performs interpretable, structure-aware, and constraint-consistent retrieval and generation. 
During the demonstration, participants can input medical or encyclopedic multi-constraint queries, visualize how the system parses constraints, performs structural matching, and generates answers, thereby experiencing an end-to-end, interactive, and explainable RAG pipeline. 
A demo video is available at \url{https://youtu.be/J8kahzmAnu0}.
\end{abstract}

\section{Introduction}
\begin{figure}[!t]
\includegraphics[width=0.99\columnwidth]{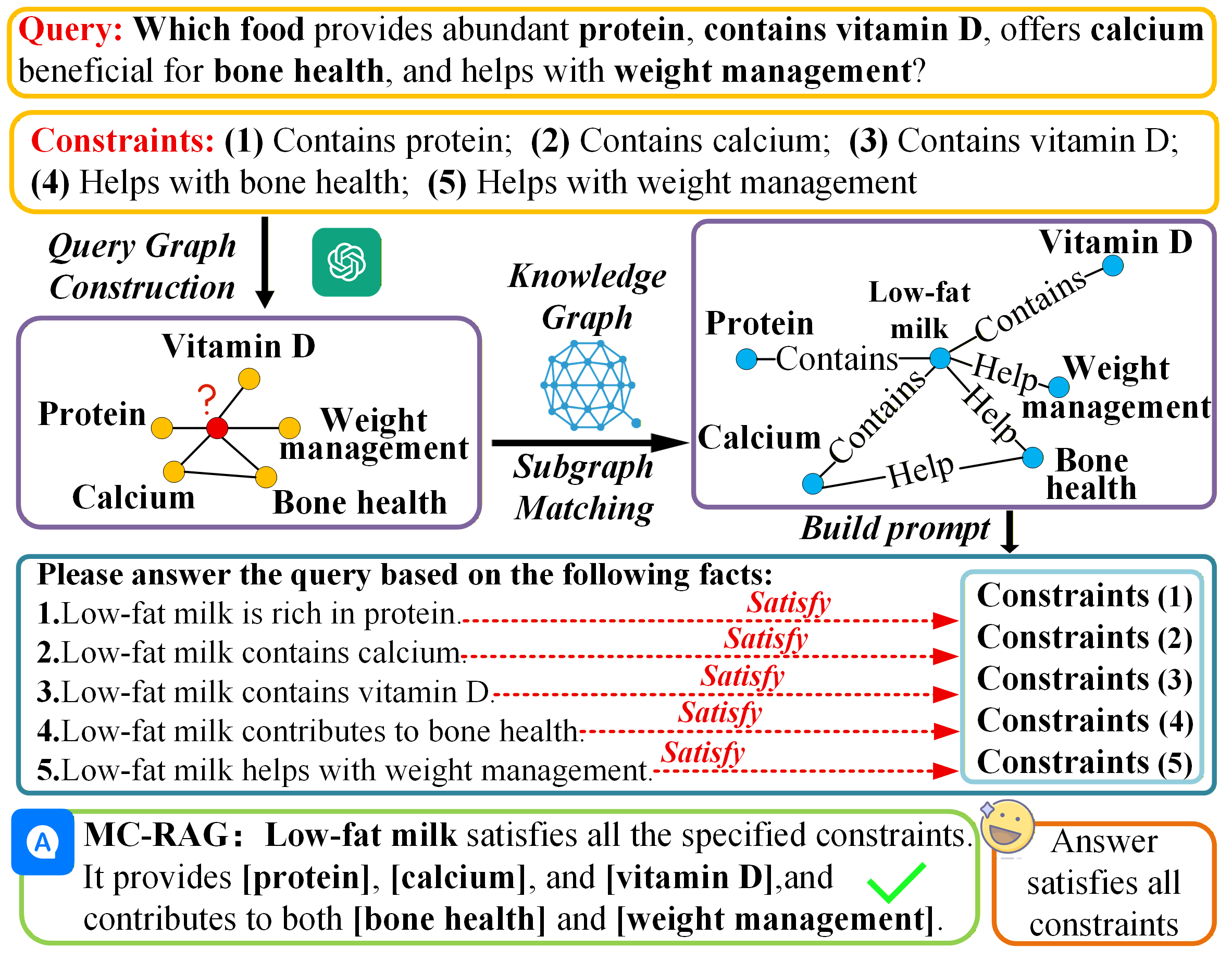}
\caption{Multi-constraint query example.}
\label{fig1}
\end{figure}

Retrieval-Augmented Generation (RAG)~\cite{b1} has become a core framework in knowledge-driven question answering systems~\cite{izacard2021leveraging,borgeaud2022improving}, where external documents are retrieved to ground the generation of LLMs. While effective for simple, factual queries, real-world user queries often impose multiple constraints to be satisfied simultaneously, rather than matched by loosely related texts to provide rough knowledge. For example, in the nutritional scenario shown in Fig.\ref{fig1}, a user may ask:
\textit{``Which food provides abundant protein, contains vitamin D, offers calcium beneficial for bone health, and helps with weight management?''}. The answer must jointly satisfy five constraints (i.e.
containing protein, calcium, and vitamin D). However, existing RAG systems struggle to address such multi-constraint requirements. Mainstream RAG systems (i.e., chunk-based RAG systems\cite{b2}) retrieve documents solely by semantic similarity between the query and text chunks, ignoring the constraints in queries~\cite{karpukhin2020dense,xu2024unsupervised,khattab2020colbert}. Recent advancements in graph-based RAG systems\cite{b3,b7,b10,he2024g} aim to mitigate the problem by organizing external knowledge as a knowledge graph and retrieves entities or subgraphs most similar to the query. Yet, such methods still rely on similarity ranking over local graph structures, and thus cannot ensure precise satisfaction of single or multiple constraints.

% To address this limitation, we propose \textsc{MC-RAG}, a
% demonstration system that integrates constraint parsing, structural
% subgraph retrieval, constraint verification, and generation grounded in
% retrieved evidence into an interactive RAG workflow.
To address this limitation, we propose \textsc{MC-RAG}, a
demonstration system built upon our structure guided RAG
framework~\cite{SG-RAG}. It integrates constraint parsing,
structural subgraph retrieval, constraint verification, and generation
grounded in retrieved evidence into an interactive RAG workflow.
% the first framework, to our knowledge, that precisely satisfies all constraints in user queries.
Our core idea is to replace similarity ranking with \emph{constraint-satisfying subgraph matching} over a knowledge graph~\cite{han2013turboiso,kim2018turboflux,Panda-system,panda,zhang2024comprehensive}, and use the matched subgraph as verifiable evidence to guide generation.
Specifically, the system first parses the natural language query into
a query graph, where the constraints are represented as nodes or edges.
Unlike prior path-dominance indexing for fully labeled query
graphs~\cite{b5}, MC-RAG handles unknown
nodes by completing missing labels during online matching. Then, it
retrieves subgraphs in the knowledge graph that are isomorphic or
approximately isomorphic to the query graph, which
satisfies all constraints. Finally, the retrieved structured evidence is
used to guide the generation.
% Specifically, the system first parses the natural language query into a query graph, where the constraints are represented as nodes or edges.
% Then, it retrieves subgraphs in the knowledge graph that are isomorphic or approximately isomorphic to the query graph~\cite{b5}, whose information satisfies all constraints.
% Finally, the retrieved structured evidence is used to guide the generation model. 

In addition, \textsc{MC-RAG} introduce an R*-Tree index to organize external knowledge represented by embeddings, which avoids the NP-complete bottleneck\cite{b4} of subgraph isomorphism on large-scale knowledge graphs and achieves near-linear retrieval efficiency.
In this demonstration, we present an interactive prototype of \textsc{MC-RAG}. 
Unlike conventional ``black-box'' QA systems, users can visually inspect the full pipeline, including query parsing, subgraph matching, and answer generation. 
The demonstration highlights three key capabilities: 
(1) \textbf{Exact structural retrieval}: subgraph matching for multi-constraint queries, enabled by path-level embeddings and R*-Tree indexing; 
(2) \textbf{Constraint-consistent generation}: generating answers only when all query constraints are satisfied; 
and (3) \textbf{Interactive interpretability}: visualizing the complete process from query graph construction to final generation.
As shown in Fig~\ref{fig1}, \textsc{MC-RAG} identifies an isomorphic subgraph centered on ``low-fat milk'' 
that satisfies all constraints, thereby producing a logically complete answer. 

In summary, \textsc{MC-RAG} reformulates multi-constraint queries as subgraph matching and enables efficient retrieval via indexing with embeddings.
The demo lets users visually inspect how constraints are parsed, matched, and verified before generation, turning RAG from a black-box process into an transparent workflow and
providing a reusable basis for future RAG systems for constraint handling.

\section{System Architecture}
\begin{figure}[!b]
\includegraphics[width=0.95\columnwidth]{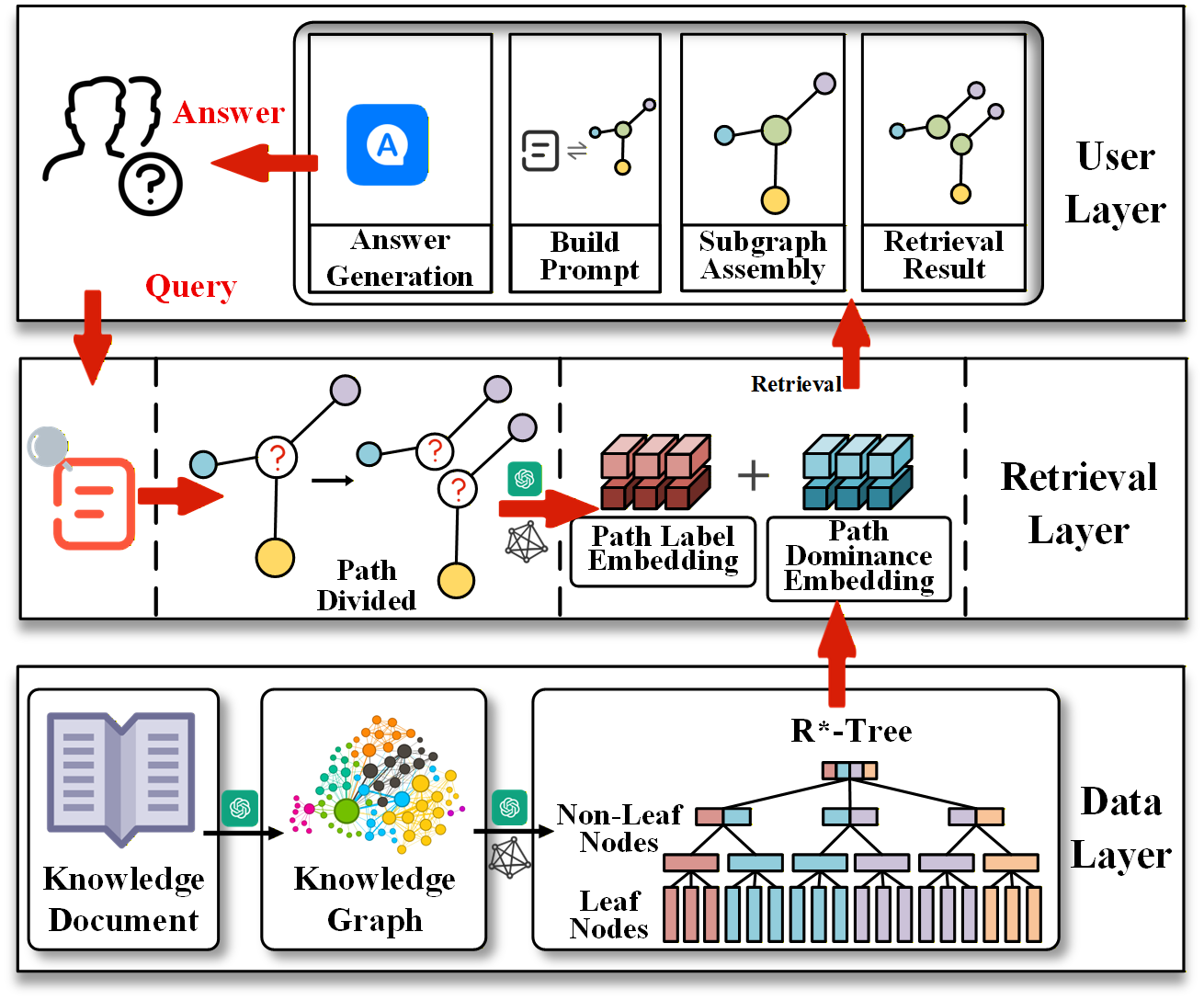}
\caption{Overall architecture of the MC-RAG system.}
\label{fig2}
\end{figure}

The overall architecture of the \textsc{MC-RAG} is shown in Fig.\ref{fig2}. 

\paragraph{Data Layer.}

To better represent the semantic and structural information of external knowledge and lay the foundation for precise subgraph matching, we design the Data Layer to extract entity and relationship from raw knowledge, construct a knowledge graph, and build efficient path-level indexes.
\textsc{MC-RAG} adopts a dual-embedding and indexing mechanism, where each path is encoded into two embeddings:
(1) a label embedding capturing the semantic meaning of nodes in the path, and
(2) a dominant embedding learned by a graph neural network (GNN) to represent the structural pattern of the path.
All path embeddings are organized in an R*-Tree.

\paragraph{Retrieval Layer.}

Retrieval Layer serves as the core component of \textsc{MC-RAG}, responsible for parsing user queries and performing subgraph matching-based retrieval on the knowledge graph. 
When a natural language question is submitted, the system first converts it into a query graph. 
The query graph is then decomposed into several paths, and the system executes path-level matching based on the R*-Tree index. 
Finally, the matched paths are aggregated and validated, 
producing a candidate subgraph that satisfies the full query conditions. 
This structured subgraph is then passed to the generation module to produce the answer.

\paragraph{User Layer.}

The User Layer is an interactive visualization interface where users can input natural language queries and observe the entire process of query parsing, subgraph retrieval, and answer generation in real time. 
This interaction mode allows users to intuitively understand how \textsc{MC-RAG} conducts interpretable and structure-driven retrieval-augmented generation.

\section{Index Construction and Subgraph Matching-Based RAG}

This section introduces the core process of \textsc{MC-RAG}.

\paragraph{Offline: Index Construction.}
We build a reusable path index over a knowledge graph to support efficient multi-constraint structural retrieval. Given raw documents, the system (i) extracts entities and relations to form a KG, and encodes each node label with an LLM to obtain semantic label embeddings; (ii) trains a lightweight GNN encoder to obtain structure-aware \emph{dominant} embeddings that are comparable across local and global neighborhoods; and (iii) enumerates KG paths of a fixed length and stores each path with a pair of \textbf{(semantic, structural)} embeddings in an R$^\ast$-Tree. The R$^\ast$-Tree organizes embedding regions for hierarchical pruning, enabling fast candidate path retrieval for downstream query matching.

\paragraph{Online: Subgraph Matching-Based RAG.}
Given a user queries, the system uses an LLM to parse constraints and construct a query graph, and normalizes query mentions to KG entities via embedding-based retrieval. The query graph is decomposed into fixed-length paths, where missing labels are handled by wildcard completion using the index to propose candidate types/entities. We then perform path-level retrieval on the R$^\ast$-Tree with joint semantic-and-structural filtering, aggregate matched paths, and validate them to form a candidate subgraph that satisfies all constraints, which is finally fed to the LLM as structured evidence for answer generation: If exact isomorphic matches exist, \textsc{MC-RAG} uses all matched subgraphs as
evidence. Otherwise, it selects the approximate match with the minimum
edit distance; if no reliable approximate match exists, it falls back to
the core entity and its one hop neighbors.

\begin{figure}[!t]
\includegraphics[width=1\columnwidth]{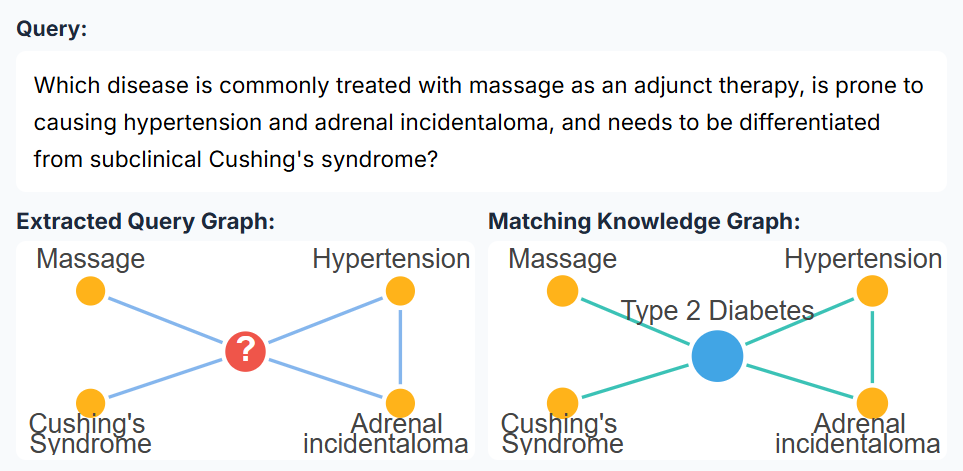}
\caption{Visualization of query graph and matched subgraphs.}
\label{fig5}
\end{figure}

\section{Demonstration Interface and Functions}
\paragraph{System Interface Overview.}

The main interface consists of three functional modules: \textit{(1) Knowledge Graph Construction}, \textit{(2) Knowledge Graph Display}, and \textit{(3) Interactive QA}.

In \textit{Knowledge Graph Construction} module, 
users can upload raw knowledge documents (PDF, DOCX, TXT, etc.), 
and the system automatically extracts entities and relations to construct a knowledge graph while displaying the processing progress in real time. 
Once construction is complete, the system generates both label and dominant embeddings for all paths in the background 
and builds an R*\mbox{-}Tree index to support subsequent structure\mbox{-}aware retrieval.
\begin{figure}[!b]
\includegraphics[width=0.95\columnwidth]{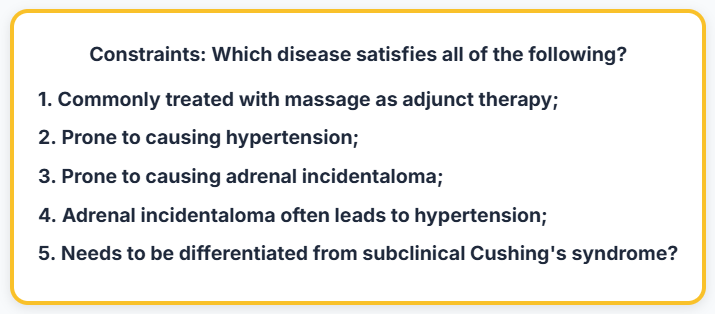}
\caption{Query parsing and constraint extraction}
\label{fig4}
\end{figure}
In \textit{Knowledge Graph Display} module, 
users can view the global topology and type distribution of the constructed graph. 
The interface supports filtering by node type, zooming, and drag\mbox{-}and\mbox{-}drop interaction. 
Different colors represent different entity types. 
The system dynamically displays the number of nodes and edges, 
helping users inspect extraction quality and graph connectivity before running queries.

In \textit{Interactive QA} module, as shown in Fig.\ref{fig5},
users can enter natural language queries such as:  
\textit{``Which disease commonly uses massage as an adjuvant therapy, is prone to have adrenal incidentaloma and hypertension, and requires differential diagnosis from subclinical Cushing's syndrome, adrenal incidentaloma often lead to hypertension?''}  
The system automatically parses the question using an LLM-based query parser to extract both semantic and structural constraints, and presents them as a structured list, as shown in Fig.\ref{fig4}.  
These constraints, together with the target entity type, form the query graph $q$.
Based on the parsed query graph, the system performs path-level matching against the indexed knowledge graph. Interactive QA module then visualizes the query graph and the retrieved candidate subgraphs that satisfy all constraints.  

\begin{figure}[!t]
\includegraphics[width=1\columnwidth]{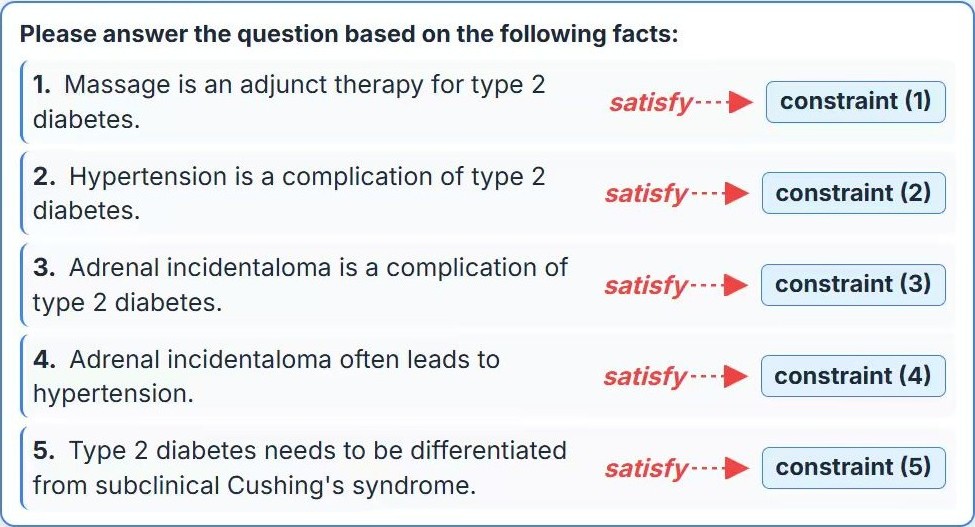}
\caption{Constraint verification before answer generation.}
\label{fig6}
\end{figure}

\begin{figure}[!t]
\includegraphics[width=1\columnwidth]{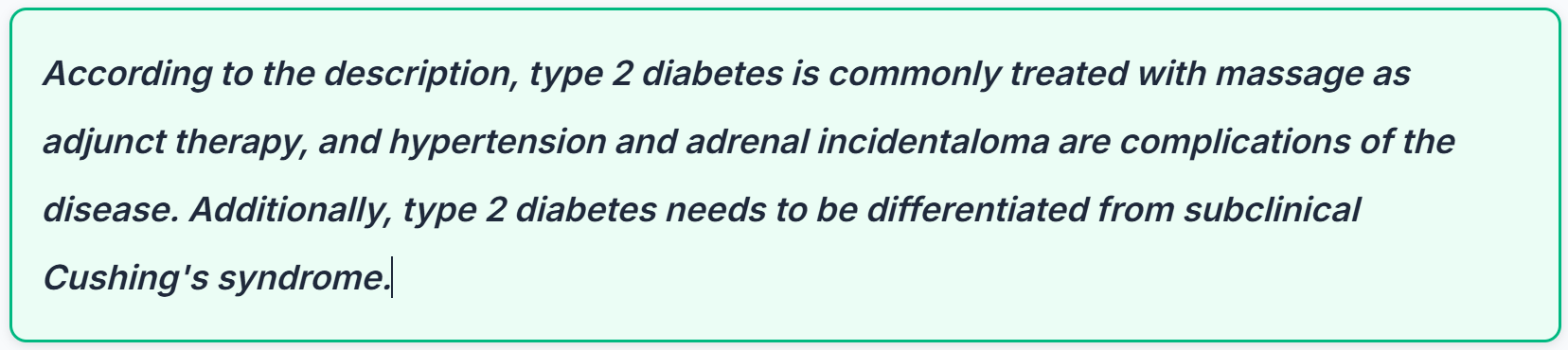}
\caption{Final answer and reasoning chain visualization.}
\label{fig7}
\end{figure}

As shown in Fig.\ref{fig5}, users can clearly observe the structural evidence underlying the generated answer.
As shown in Fig.~\ref{fig6}, 
once all constraints are verified, the generation module converts the final matched subgraph into a structured prompt  and invokes the language model to produce a natural language explanation. 
As shown in Fig.~\ref{fig7}, 
the interface simultaneously displays the generated answer and its logical reasoning chain.
\paragraph{Experimental Evaluation.}
Before the demo, we systematically evaluated \textsc{MC-RAG} using the multi-constraint dataset \textit{ERQA}\footnote{ERQA:(\url{https://github.com/CAU-X-AI-Lab/ERQA})} and the single-constraint dataset \textit{Natural Questions}\cite{b9}.
\textit{ERQA} covers $20$ domains and contains approximately $206,000$ queries, with each query involving an average of $4.6$ constraints.
Experimental results show that \textsc{MC-RAG} significantly outperforms representative state-of-the-art RAG systems and LLMs.
Compared with NaiveRAG\cite{b2}, GraphRAG\cite{b10}, LightRAG\cite{b3}, KAG\cite{b7}, GPT-$5$, GLM-4V, Gemini-$2.5$ and Qwen-$3$, \textsc{MC-RAG} achieves consistent absolute gains on precision, recall, F1 and Hit@1 (up to $+37.04$ points in Hit@1, $+42.45$ points in Recall, and $+0.40$ points in F1).
In terms of efficiency, \textsc{MC-RAG} achieves an average end-to-end generation time of $10.8$ seconds per query; the subgraph matching pipeline takes about $6–7$ seconds, comparable to KAG.
During the offline index construction phase, \textsc{MC-RAG} requires approximately $11$ hours in total, slightly longer than LightRAG(around 10 hours).
The index can be reused across datasets, and the GNN encoder is trained only once offline.

\section*{Acknowledgements}
This work was supported by the China Agricultural University ``Young Researcher'' Start-up Fund No.~QNYJY2024144 and the Visiting Scholar Program of the China Scholarship Council (CSC) No.~202506350123.
\section*{Contribution Statement}
Xiao Zhang and Yang Wan contributed equally to this work.
Miao Xie is the corresponding author.

\bibliographystyle{named}
\bibliography{ijcai26}

\end{document}